\documentclass[runningheads,a4paper]{llncs}

\usepackage{amssymb}
\usepackage{graphicx}

\usepackage{url}
\urldef{\mailsa}\path|{alfred.hofmann, ursula.barth, ingrid.haas, frank.holzwarth,|
\urldef{\mailsb}\path|anna.kramer, leonie.kunz, christine.reiss, nicole.sator,|
\urldef{\mailsc}\path|erika.siebert-cole, peter.strasser, lncs}@springer.com|    
\newcommand{\keywords}[1]{\par\addvspace\baselineskip
\noindent\keywordname\enspace\ignorespaces#1}

\begin{document}

\hyphenation{wiki-space}

\mainmatter  

\title{Collaborative Structuring of Knowledge\\ by Experts and the Public}

\titlerunning{Collaborative Structuring of Knowledge by Experts and the Public}

%
%
\author{Tom Morris\inst{1}%
\and Daniel Mietchen\inst{2}}
\authorrunning{Tom Morris and Daniel Mietchen}

\institute{\url{http://www.citizendium.org/User:Tom_Morris}
\and
\url{http://www.citizendium.org/User:Daniel_Mietchen}\\Correspondence: Daniel.Mietchen (at) uni-jena (dot) de }


%
%

\toctitle{Collaborative Structuring of Knowledge by Experts and the Public}
\tocauthor{Tom Morris and Daniel Mietchen}
\maketitle

\begin{abstract}
There is much debate on how public participation and expertise can be brought together in collaborative knowledge environments. One of the experiments addressing the issue directly is Citizendium. In seeking to harvest the strengths (and avoiding the major pitfalls) of both user-generated wiki projects and traditional expert-approved reference works, it is a wiki to which anybody can contribute using their real names, while those with specific expertise are given a special role in assessing the quality of content. Upon fulfillment of a set of criteria like factual and linguistic accuracy, lack of bias, and readability by non-specialists, these entries are forked into two versions: a stable (and thus citable) approved "cluster" (an article with subpages providing supplementary information) and a draft version, the latter to allow for further development and updates. We provide an overview of how Citizendium is structured and what it offers to the open knowledge communities, particularly to those engaged in education and research. Special attention will be paid to the structures and processes put in place to provide for transparent governance, to encourage collaboration, to resolve disputes in a civil manner and by taking into account expert opinions, and to facilitate navigation of the site and contextualization of its contents.
\keywords{open knowledge, open education, open science, open \\governance, wikis, expertise, Citizendium, Semantic Web}
\end{abstract}

\section{Introduction}

\begin{quote}
{\it Science is already a wiki if you look at it a certain way. It's just a highly inefficient one -- the incremental edits are made in papers instead of wikispace, and significant effort is expended to recapitulate existing knowledge in a paper in order to support the one to three new assertions made in any one paper.  } \begin{flushright} John Wilbanks \cite{Wilbanks2009}
\end{flushright}\end{quote}

There are many ways to structure knowledge, including collaborative arrangements of digital documents. Only a limited number of the latter ones have so far been employed on a larger scale. Amongst them are wikis~-- online platforms which allow the aggregation, interlinking and updation of diverse sets of knowledge in an Open Access manner, i.e. with no costs to the reader. 

\subsection{Wikis as an example of public knowledge environments online}

As implied by the introductory quote, it is probably fair to say that turning science (or any system of knowledge production, for that matter) into a wiki (or a set of interlinked collaborative platforms) would make research, teaching and outreach much more transparent, less prone to hype, and more efficient. Just imagine you had a time slider and could watch the history of research on general relativity, plate tectonics, self-replication, or cell division unfold from the earliest ideas of their earliest proponents (and opponents) onwards up to you, your colleagues, and those with whom you compete for grants. So why don't we do it?

Traditionally, given the scope of a particular journal, knowledge about specialist terms (which may describe completely non-congruent concepts in different fields), methodologies, notations, mainstream opinions, trends, or major controversies could reasonably be expected to be widespread amongst the audience, which reduced the need to redundantly say and then repeat the same things all over again and again (in cross-disciplinary environments, there is a higher demand for proper disambiguation of the various meanings of a term). Nonetheless, redundancy is still quite visible in journal articles, especially in the introduction, methods, and discussion sections and the abstracts, often in a way characteristic of the authors (such that services like eTBLAST and JANE can make qualified guesses on authors of a particular piece of text, with good results if some of the authors have a lot of papers in the respective database, mainly PubMed, and if they have not changed their individual research scope too often in between).

A manuscript well-adapted to the scope of one particular journal is often not very intelligible to someone outside its intended audience, which hampers cross-fertilization with other research fields (we will get back to this below). When using paper as the sole medium of communication there is not much to be done about this limitation. Indeed, we have become so used to it that some do not perceive it as a limitation at all. Similar thoughts apply to manuscript formatting. However, the times when paper alone reigned over scholarly communication have certainly passed, and wiki-like platforms provide for simple and efficient means of storing information, updating it and embedding it into a wider context.

Cross-field fertilization, for example, is crucial with respect to interdisciplinary research projects, digital libraries and multi-journal (or indeed cross-disciplinary) bibliographic search engines (e.g. Google Scholar), since these dramatically increase the likelihood of, say, a biologist stumbling upon a not primarily biological source relevant to her research (think shape quantification or growth curves, for instance). What options do we have to systematically integrate such cross-disciplinary hidden treasures with the traditional intra-disciplinary background knowledge and with new insights resulting from research?

The by now classical example of a wiki environment are the Wikipedias, a set of interlinked wikis in multiple languages where basically anyone can edit any page, regardless of subject matter expertise or command of the respective language. As a consequence of this openness, the larger Wikipedias have a serious problem with vandalism: take an article of your choice and look at its history page for reverts - most of them will be about neutralizing subtle or blunt forms of destructive edits that do nothing to improve the quality of the articles, but may reduce it considerably. Few of these malicious edits persist for long  \cite{Priedhorsky:2007}, but finding and fixing them takes time that could better be spent on improving articles. This is less of an issue with more popular topics for which large numbers of volunteers may be available to correct "spammy" entries but it is probably fair to assume that most researchers value their time too much to spend it on repeatedly correcting information that had already been correctly entered. Other problems with covering scientific topics at the Wikipedias include the nebulous notability criteria which have to be fulfilled to avoid an article being deleted, and the rejection of "original research" in the sense of not having been peer reviewed before publication. Despite these problems, one scientific journal~-- RNA Biology~-- already requires an introductory Wikipedia article for a subset of papers it is to publish \cite{RNABiol}.

Peer review is indeed a central aspect of scholarly communication, as it paves the way towards the reproducibility that forms one of the foundations of modern science. Yet we know of no compelling reason to believe that it works better before than after the content concerned has been made public (doing it beforehand was just a practical decision in times when journal space was measured in paper pages), while emerging movements like Open Notebook Science~-- where claims are linked directly to the underlying data that are being made public as they arise~-- represent an experiment in this direction whose initial results look promising and call into question Wikipedia's "no original research" as a valid principle for generating encyclopaedic content.

Although quite prominent at the moment, the Wikipedias are not the only wikis around, and amongst the more scholarly inclined alternatives, there are even a number of wiki-based journals, though usually with a very narrow scope and/or a low number of articles. On the other hand, Scholarpedia (which has classical peer review and an ISSN and may thus be counted as a wiki journal, too \cite{Scholarpedia}), OpenWetWare \cite{OpenWetWare}, Citizendium \cite{Citizendium} and the Wikiversities \cite{Wikiversity} are cross-disciplinary and structured (and of a size, for the moment) such that vandalism and notability are not really a problem. With minor exceptions, real names are required at the first three, and anybody can contribute to entries about anything, particularly in their fields of expertise. None of these is even close to providing the vast amount of context existing in the English Wikipedia but the difference is much less dramatic if the latter were broken down to scholarly useful content. Out of these four wikis, only OpenWetWare is explicitly designed to harbour original research, while the others allow different amounts thereof. Furthermore, a growing number of yet more specialized scholarly wikis exist (e.g. WikiGenes \cite{WikiGenes}, the Encyclopedia of Earth \cite{EoEarth}, the Encyclopedia of Cosmos \cite{EoCosmos}, the Dispersive PDE Wiki \cite{Dispersive-PDE}, or the Polymath Wiki \cite{Polymath}), which can teach us about the usefulness of wikis within specific academic fields. 

\section{The Citizendium model of wiki-based collaboration}
Despite the above-mentioned tensions between public participation and expertise in the collaborative structuring of knowledge, it is not unreasonable to expect that these can be overcome by suitably designed public knowledge environments, much like Citizen Science projects involve the public in the generation of scientific data. One approach at such a design is represented by Citizendium. The founder of Citizendium~-- Larry Sanger~-- is the co-founder of Wikipedia. The two projects share the common goal of providing free knowledge to the public, they are based on variants of the same software platform, and they use the same Creative Commons-Attribution-Share Alike license \cite{CC-BY-SA}. Yet they differ in a number of important ways, such that Citizendium can be seen as composed of a Wikipedia core (stripped down in terms of content, templates, categories and policies), with elements added that are characteristic of the other wiki environments introduced above: A review process leading to stable versions (as at Scholarpedia), an open education environment (as at Wikiversity) and an open research environment (as at OpenWetWare). Nonetheless, assuming that the reader is less familiar with these three latter environments, we will follow previous commenters and frame the discussion of Citizendium in terms of properties differentiating it from Wikipedia, and specifically the latter's English language branch \cite{Wikipedia:En}. 

\subsection{Real names}
The first of these is simply an insistence on real names. While unusual from a Wikipedia perspective, this is custom in professional environments, including traditional academic publishing and some of the above-mentioned wikis, e.g. Scholarpedia and Encyclopedia of Earth. It certainly excludes a number of legitimate contributors who prefer to remain anonymous but otherwise gives participants accountability and allows to bring in external reputation to the project. 

\subsection{Expert guidance}

\begin{figure}
\centering
\includegraphics[width=12.2cm]{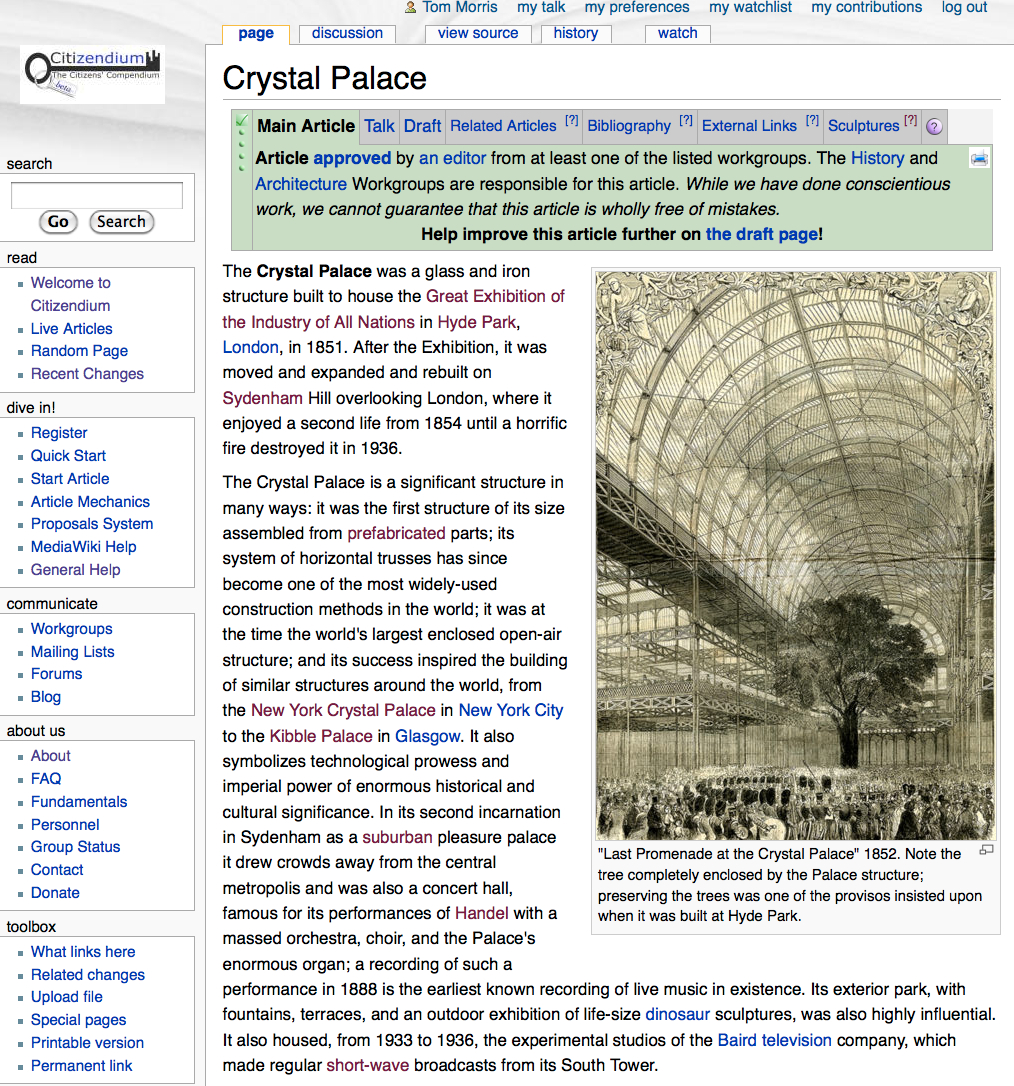}
\caption{Screenshot of the main page of the [[Crystal Palace]] cluster while logged in using a monobook skin that is the default at Wikipedia. It shows the \textit{green cluster bar} that indicates that the page has been approved and links to all the supages. Also visible is the \textit{status indicator} (green dots on the left topped by green tick), mention of \textit{"an editor"} to distinguish the number of editors involved (some pages can be approved by one rather than three editors), links to the workgroups which have approval rights for the article (in this case: the \textit{History and Architecture Workgroups}), a prominent \textit{disclaimer} (unapproved articles have a much strong disclaimer), and links to the 'unstable' \textit{draft} version of the article which any registered contributor can update. Like traditional encyclopaedic environments, Citizendium does not require every statement to be referenced, in the interest of narrative flow.}
\label{fig:Approved}
\end{figure}
To compose and develop articles and to embed them in the multimedial context of a digital knowledge environment, expert guidance is important. Of course, many experts contribute to Wikipedia, and the Wikipedias in turn have long started to actively seek out expert involvement, yet the possibility to see their edits overturned by anonymous users that may lack even the most basic education in that field keeps professionals away from spending their precious time on such a project. The Citizendium approach of verifying expertise takes a different approach~-- sometimes termed "credentialism"~-- that rests on a common sense belief that some people do know more than others: it is sometimes the case that the thirteen-year-old kid in Nebraska does know more than the physics professor. But most of the time, at least when matters of physics are concerned, this is not the case. The role the experts have at Citizendium is not, as frequently stated in external comments, that of a supreme leader who is allowed to exercise his will on the populace. On the contrary, it is much more about guiding. We use the analogy of a village elder wandering around the busy marketplace \cite{Basar} who can resolve disputes and whom people respect for their mature judgement, expertise and sage advice. Wikipedia rejects "credentialism" in much the same way that the Internet Engineering Task Force (IETF) does. David Clark summarised the IETF process thusly \cite{Clark:1992}: "We reject kings, presidents and voting. We believe in rough consensus and running code." In an open source project, or an IETF standardisation project, one can decide a great many of the disputes with reference to technical reality: the compiler, the existing network protocols etc. If the code doesn't compile, think again. For rough consensus to happen under such circumstances, one needs to get the people together who have some clear aim in mind: getting two different servers to communicate with one another. The rough consensus required for producing an encyclopaedia article is different~-- it should attempt to put forward what is known, and people disagree on this to a higher degree than computers do on whether a proper connection has been established. It is difficult to get "rough consensus, running code" when two parties are working on completely different epistemological standards. At this point, one needs the advice of the village elderly who will vet existing content and provide feedback on how it can be expanded or otherwise improved. Upon fulfillment of a set of criteria like factual and linguistic accuracy, lack of bias, and readability by non-specialists, these vetted entries are forked into two versions: a stable (and thus citable) approved "cluster" (an article with subpages providing supplementary information) and a draft version, the latter to allow for further development and updates (cf.\ Fig.\ \ref{fig:Approved}).

The respect for experts because of their knowledge of facts is only part of the reasoning: the experts point out and correct factual mistakes, but they also help to guide the structuring of content within an article and by means of the subpages. The experts bring with them the experience and knowledge of years of in-depth involvement with their subject matter, and the project is designed to make best use of this precious resource, while still allowing everyone to participate in the process. Of course, experts are likewise free to bring in content, be it within their specialty or in other areas, where others take over the guiding role. The Citizendium can also host 'Signed Articles', which are placed in a subpage alongside the main article. A Signed Article is an article on the topic described by a recognised expert in the field, but can express opinions and biases in a way that the main article ought not to.

\begin{figure}
\centering
\includegraphics[width=11.2cm]{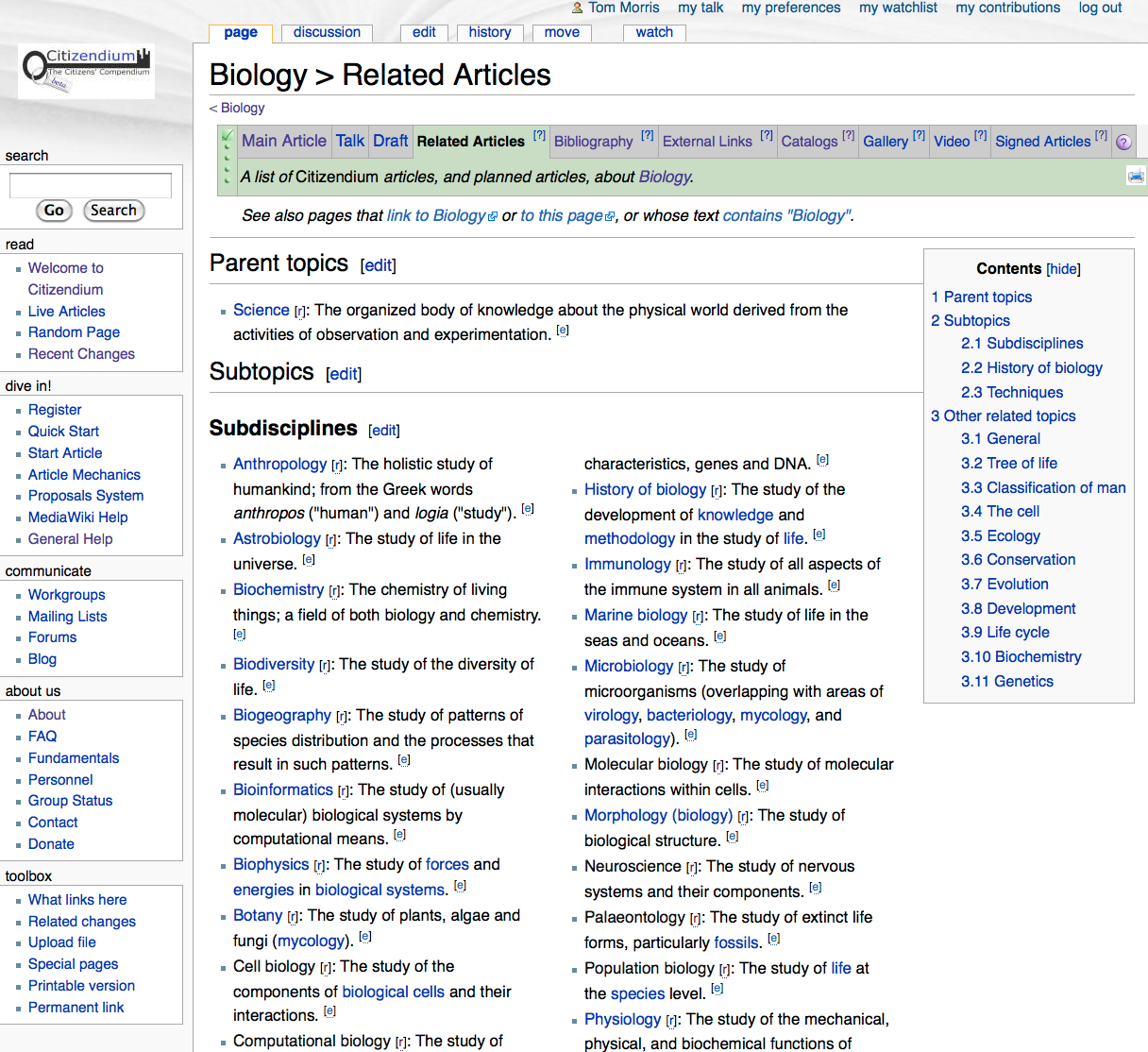}
\includegraphics[width=11.2cm]{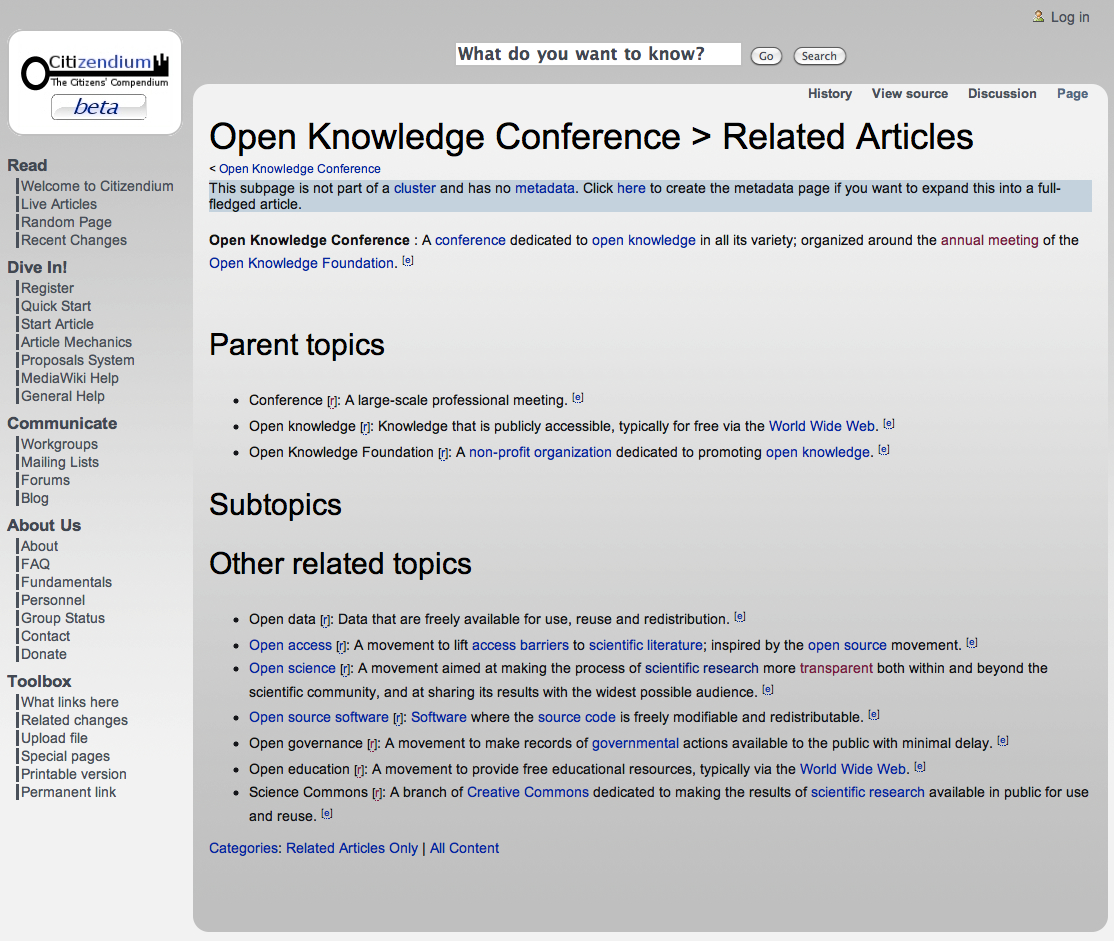}
\caption{{\bf Top:} Screenshot of the Related Articles subpage from the [[Biology]] cluster (which is approved) while logged in. It shows the \textit{Parent topics} and the first section of the \textit{Subtopics}~-- \textit{subdisciplines}. For each related article, there is a short definition or description of the topic, and a link to its Related Articles subpage (hidden behind the \textit{[r]}), as well as instructions on mouseover and a Table of \textit{Content}. {\bf Bottom:} Related Articles subpage from the [[Open Knowledge Conference]] cluster (while logged out) which has not yet been converted to subpage style but can already be used for structuring information related to the topic. In principle, on could also think of adding [[Open Knowledge Conference 2010]] as a subtopic and using this article for conference blogging. However, the current MediaWiki software cannot handle parallel editing by multiple users, though tools like Etherpad \cite{Etherpad} have shown that it is feasible.}
\label{fig:Related-Approved}
\end{figure}

\subsection{Contextualization}
Citizendium attempts to structure knowledge in a different way. Each article on Citizendium can make comprehensive use of Subpages, i.e. pages providing additional information that are subordinate to an article's page. Some of these~-- e.g. the Sculptures subpage in Fig.\ \ref{fig:Approved}~-- are similar to but more flexible than the supplementary online materials now being published routinely along scholarly articles. Two subpages types are different, with keywords  and running title being the closest analogues from academic papers: All pages are encouraged to have a short Definition subpage (around 30 words or 150 characters) which defines or describes the subject of the page. They are also encouraged to have a comprehensive Related Articles subpage, which uses templates to pull in the definitions from the pages that it links to (a feature that relies on the absence of vandalism). If one looks at the Related Articles subpage of [[Biology]] (cf.\ Fig.\ \ref{fig:Related-Approved}, top), one can see the parent topics of biology (science), the subtopics - subdisciplines of biology like zoology, genetics and biochemistry, articles on the history of biology and techniques used by biologists - and finally other related topics, including material on the life cycle, the various biochemical substances like DNA and proteins, the components of the cell, and other specialised language. This Related Articles page gives a pretty comprehensive contextual introduction to what biology is all about, and is structured by the authors of the article in a way that is consistent across the site (cf.\ Fig.\ \ref{fig:Related-Approved}, bottom). This goes beyond Wikipedias categories, "See also" sections and ad-hoc infoboxes. Citizendium's approach can be considered as an exploratory next step towards linking encyclopaedic content with the Semantic Web. 

\begin{figure}
\centering
\includegraphics[width=12.2cm]{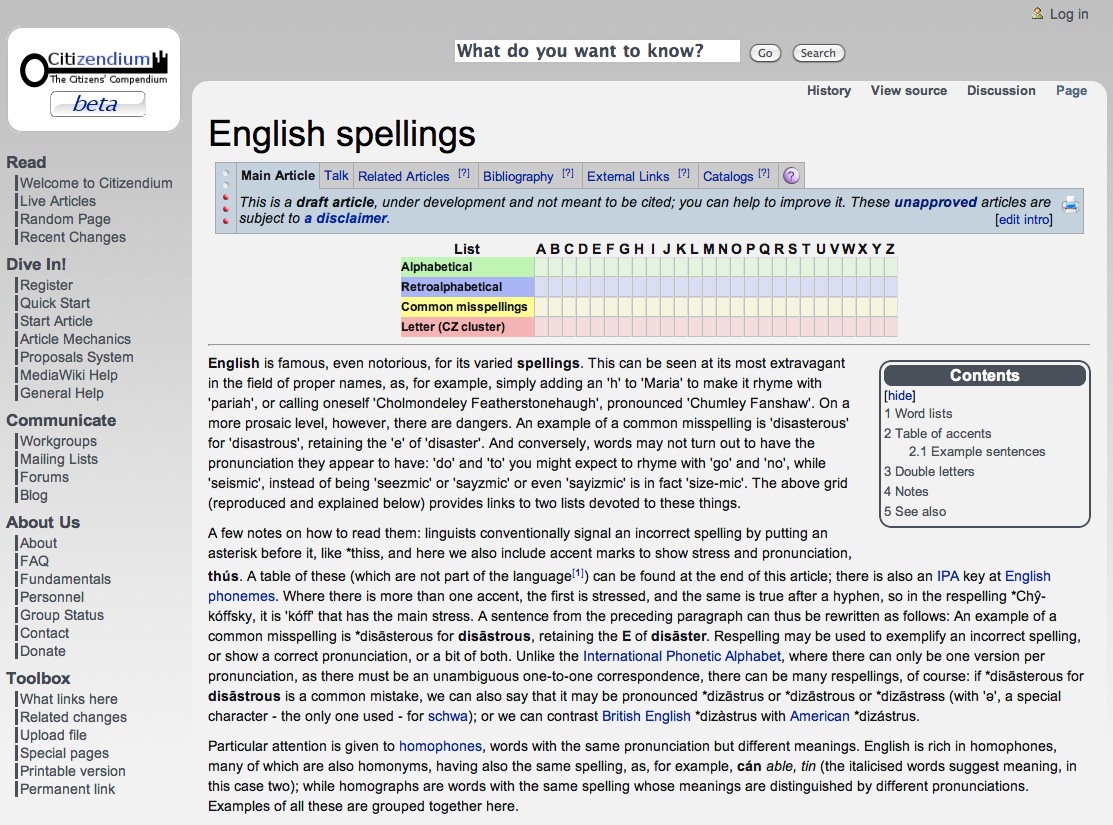}
\includegraphics[width=12.2cm]{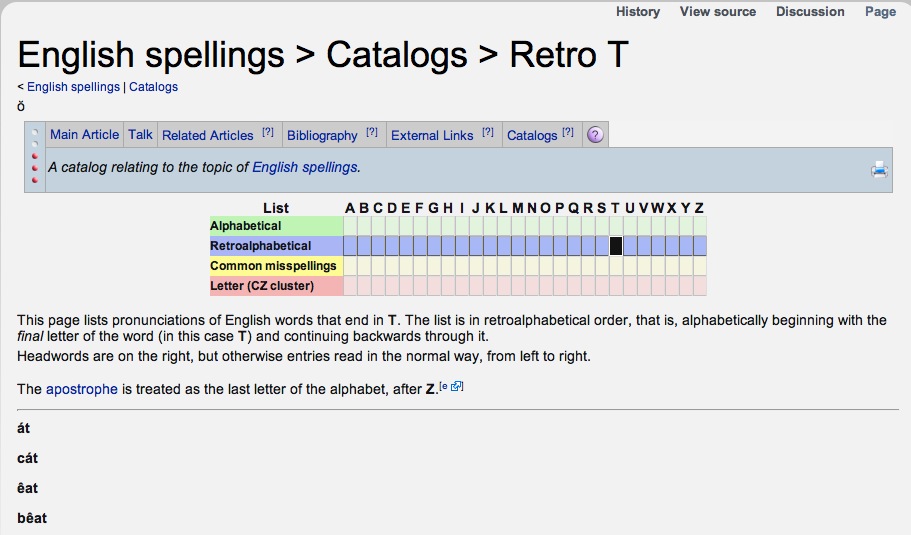}
\caption{{\bf Top:} Screenshot of the main page of the [[English spellings]] cluster while logged out. It shows the \textit{blue cluster bar} that indicates that the page has not been approved and links to  all the supages. Also visible is the \textit{status indicator} (red dots on the left topped by grey dots, indicating the status of a 'developing' article), and a stronger \textit{disclaimer} than on approved pages. Below this standard header, a set of templates links to 'Catalog' subpages that collect links, for each letter of the English alphabet, to \textit{Alphabetical} and \textit{Retroalphabetical} lists of spellings, to lists of \textit{Common misspellings} as well as to an article on the specific \textit{Letter}. {\bf Bottom:} Close-up of the Catalogs subpage hosting the retroalphabetical list of English spellings for the letter T, again cross-linked with all the other subpages in that cluster.}
\label{fig:Spellings}
\end{figure}

Subpages (a further usage example is in (cf.\ Fig.\ \ref{fig:Spellings}) are one way in which Citizendium is attempting to go beyond what is provided in either traditional paper-based encyclopaedias or by Wikipedia: to engage with context, with related forms of knowledge, and to emancipate knowledge from the page format to which it was confined in the print era. Marx wrote that "Philosophers have hitherto only interpreted the world in various ways; the point is to change it" \cite{Marx}. Traditional encyclopaedias attempt to reflect the world, but we are attempting to go further. The open science movement - which has formed around the combination of providing open access to journal articles, making scientific data more openly available in raw forms, using and sharing open source software and experimenting with some of the new techniques appearing from the community that is formed under the 'Web 2.0' banner - is exploring the edge of what is now possible for scientists to do to create new knowledge. Some of the electronic engagements by academics has been for actual research benefit, some has just been PR for universities - doing podcasts to sound 'relevant'. The Citizendium model, while a little bit more traditional than some of the open science platforms, is willing to try a variety of new things. Wikipedia has produced a pretty good first version of a collaboratively written encyclopedia~-- the challenge is to see if we can go further and produce a citizens' compendium of structured and comprehensive knowledge and update it as new evidence or insights arise. 

\subsection{Open governance}
Citizendium has an evolving, but hopefully soon-to-be clearly defined governance process - a Charter is in the process of being drafted by an elected group of writers that will allow for democratic governance and oversight. The broad outline is this: we will have a democratically elected Editorial Council which will deal with content policy and resolving disputes regarding content, and we will also have a Management Committee, responsible for anything not related to content. The Management Committee appoint Constables who uphold community policy regarding behaviour. Disputes with the Constables can be brought to an Ombudsman selected by the Editorial Council and Management Committee. At the time of writing, the charter is still to be ratified by the community. One of the reasons we have this is that although there is a cost to having bureaucracy and democracy, the benefits of having an open governance process outweigh the costs. We have a real problem when governments of real-life communities are controlled by shadowy cabals who invoke byzantine legal codes - all the same problems would seem to apply to online communities. With a Wikipedia article, the debate seems to shift very quickly from the truth value or relevance of the content itself into often ritualized arguments about acronyms (AfDs, NPOV, CSD, ArbCom, OR etc.). There is always a challenge in any knowledge-based community in attempting to reconcile a fair and democratic process with a meritocratic respect for expertise. There are no easy answers - if we go too far towards bureaucracy, we risk creating a system where management is separated from the actual day-to-day writing of the site, while if we attempt to let the site 'manage itself', we risk creating a rather conservative mob rule that doesn't afford due process to interested outsiders. A more traditional management structure, combined with real names and civility, should help those outside of the online community - the many experts in real life who work in universities, in business and in public life - to participate on an equal footing. Hopefully, if we get the governance decisions right, we can also not get in the way of the people who engage on hobbyist terms with Citizendium. 

\subsection{Open education}
An important part of the governance process is collaboration with partners external to the Citizendium. One of our initiatives~-- called Eduzendium~-- provides for educators in higher education to assign work on wiki articles as part of a course. We have most recently had politics students from the Illinois State University work on articles on pressure groups in American public life, as well as medical students from Edinburgh, biologists from City University of New York and the University of Colorado at Boulder, finance students from Temple University and others. These courses reserve a batch of articles for the duration of the course, and assign each article to one or more students. The course instructor can reserve the articles for just the group of students enrolled in the course, or invite the wider Citizendium community to participate. Much of the formatting is achieved via course-specific templates that can be generated semi-automatically by the instructor and applied throughout the course pages, so that course participants can concentrate on content.
    
\section{Open questions}
The project is still young, resulting in a number of challenges and opportunities. In many fields, Citizendium does not meet our own standards~-- we do not have a full range of expert editors. Larry Sanger once envisioned that Citizendium could reach 100,000 articles by 2012. This would, on average, require about 150 new articles a day to reach; the current level is around 15. It is not obvious how the necessary shift from linear to exponential growth can be achieved.

Motivating both editors and authors to take part in both writing and approving of content remains a difficult challenge~-- most experts have very little time to offer for projects that do not contribute to the metrics according to which their performance is evaluated, and others shy away from contributing under their real name and in the presence of experts. Another problem is that the initial structure of the community, and the nature of its interaction with Wikipedia, has led to a few articles on popular pseudoscientific topics which are hard to handle from an editorial perspective because those willing to invest their time on the topics are usually heavily biased in their approach, and most of those capable of evidence-based comment prefer not to contribute to these topics.

The project also needs to allow for more feedback by non-registered readers, without harming the currently very collegial atmosphere that is to a large extent due to the real-name policy and the respect for expertise. We may need to explore how to codify our core policies and collaboration model as a possible MediaWiki extension, from which other wikis could possibly benefit~-- online, "code is law" \cite{Lessig}, as is currently being highlighted by sites like Stack Overflow which have changed the social interactions of participants by changing formal features of user experience and social structure.  We need to find financial backing and support. So far, the project has been run on a basically volunteer-only basis, yet the envisioned growth and improvement of English-language content and the possible start of branches in other languages require a higher degree of professionalisation, for which the upcoming Charter is meant as a basis.
    
\section{Open perspectives}

Citizendium is open for partnerships with other open science and online knowledge communities and projects. Possible candidate projects would include, for instance, AcaWiki \cite{AcaWiki} for references, OpenWetWare for primary research, and Open Access journals \cite{DOAJ} as possible content providers, and of course the Wikipedias and other public wikis for exchange on matters of content management, community development and user experience. The key strength we think the Citizendium model brings is a greater focus on knowledge contextualization: it will be interesting to see whether we can evolve the social model for knowledge production to keep up with changes in the technological possibilities. Many in the Citizendium community are looking forward to working alongside both academics and those working in the Semantic Web community to tie Citizendium into data projects. We feel that despite the commoditization of Web 2.0 technologies, there is still plenty of opportunities for reinventing and experimenting with new ways to render and collaborate on knowledge production and to see if we can build a more stable, sustainable and collegial atmosphere~-- with democratic and meritocratic elements~-- for experts and the public to work together.

\subsubsection*{Acknowledgments.} The authors wish to thank Russell D. Jones, Howard C. Berkowitz, Steven Mansour and Peter Schmitt for critical comments on earlier versions of this draft as well as Claudia Koltzenburg, Fran\c{c}ois Dongier and Charles van den Heuvel for helpful discussions.


\begin{thebibliography}{4}

\bibitem{AcaWiki} AcaWiki,\\ \url{http://acawiki.org/} \\All URLs referenced in this article were functional as of March 31, 2010.

\bibitem{Citizendium} Citizendium,\\ \url{http://www.citizendium.org/}

\bibitem{Clark:1992} Clark, D.: Plenary lecture, "A Cloudy Crystal Ball~-- Visions of the Future", Proc. 24th IETF: 539 (1992),\\ \url{http://www.ietf.org/proceedings/prior29/IETF24.pdf}

\bibitem{CC-BY-SA} Creative Commons-Attribution-Share Alike license 3.0,\\ \url{http://creativecommons.org/licenses/by-sa/3.0/}

\bibitem{DOAJ} Directory of Open Access Journals,\\ \url{http://www.doaj.org/}

\bibitem{Dispersive-PDE} Dispersive PDE Wiki,\\ \url{http://tosio.math.utoronto.ca/wiki/}

\bibitem{EoCosmos}  Encyclopedia of Cosmos,\\ \url{http://www.cosmosportal.org/}

\bibitem{EoEarth} Encyclopedia of Earth,\\ \url{http://www.eoearth.org}

\bibitem{Etherpad} Etherpad source code,\\ \url{http://code.google.com/p/etherpad/}

\bibitem{Lessig} Lessig, Lawrence: Code and Other Laws of Cyberspace,\\ \url{http://codev2.cc/}

\bibitem{Marx} Marx, Karl: Theses on Feuerbach,\\ \url{http://www.marxists.org/archive/marx/works/1845/theses/theses.htm}

\bibitem{OpenWetWare} OpenWetWare,\\ \url{http://www.openwetware.org/}

\bibitem{Polymath} Polymath WIki,\\ \url{http://michaelnielsen.org/polymath1/}

\bibitem{Priedhorsky:2007}
Priedhorsky R, Chen J, Lam STK, Panciera K, Terveen L, et~al. (2007) Creating,
  destroying, and restoring value in wikipedia.
In: GROUP '07: Proceedings of the 2007 international ACM conference
  on Supporting group work. New York, NY, USA: ACM, pp. 259--268.
\url{http://doi.acm.org/10.1145/1316624.1316663}.

\bibitem{Basar} Raymond, Eric S.: The Cathedral and the Bazaar, \\ \url{http://www.catb.org/~esr/writings/homesteading/}
 
\bibitem{RNABiol} RNA Biology, Guidelines for the RNA Families Track,\\ \url{http://www.landesbioscience.com/journals/rnabiology/guidelines/}

\bibitem{Scholarpedia} Scholarpedia,\\ \url{http://www.scholarpedia.org/}

\bibitem{WikiGenes} WikiGenes,\\ \url{http://www.wikigenes.org/} 

\bibitem{Wikipedia:En} English Wikipedia,\\ \url{http://en.wikipedia.org} 

\bibitem{Wikiversity} Wikiversity,\\ \url{http://www.wikiversity.org} 

\bibitem{Wilbanks2009} Wilbanks, J.: Publishing science on the web,\\ \url{http://scienceblogs.com/commonknowledge/2009/07/publishing_science_on_the_web.php}

\end{thebibliography}
\end{document}